\documentclass[%
 reprint,
 amsmath,amssymb,
 aps,
]{revtex4-2}

\usepackage{graphicx}
\usepackage{dcolumn}
\usepackage{bm}

\usepackage{xcolor}

\begin{document}

\newcommand{\ContentPath}[1]{#1}

\newcommand{\BlankLineAboveAndBelow}[1]{\phantom{.}\\ #1 \phantom{.}\\}

\newcommand{\EqRef}[1]{Eq.~(\ref{#1})}

\newcommand{\SpinHamiltonianSIAT}{{\cal{E}}_{spin}}
\newcommand{\ElectronMAE}{{\cal{E}}_{electron}}

\newcommand{\PhiTetta}{\varphi, \theta}

\newcommand{\eqdot}{\,\,\, .}
\newcommand{\eqcomma}{\,\,\, ,}

\newcommand{\synonymize}[1]{\textcolor{brown}{#1}}

\definecolor{orange}{RGB}{255,127,0}
\newcommand{\ToCite}[1]{\textcolor{orange}{[#1]}}

\newcommand{\Attention}[1]{\textcolor{red}{#1}}

\newcommand{\regexp}{\mathrm{exp}}

\newcommand{\regcos}{\mathrm{cos}}
\newcommand{\regsin}{\mathrm{sin}}

\newcommand{\energy}{E}

\newcommand{\GkE}{{\cal{G}}^{\sigma}(\energy, \bm{k})}

\newcommand{\GkEspinor}{{\cal{G}}(\energy, \bm{k})}
\newcommand{\GkhatchEspinor}{{\cal{G}}(\energy, \bm{k'})}

\newcommand{\IntImTrLSigma}[1]
{
\int_{-\infty}^{\energy_{F}}
\mathrm{Im} \,
\mathrm{Tr_{L, \sigma}}
\big[
#1
\big]
\, dE 
}

\newcommand{\Operator}{  \hat{ {\cal{F}} }  }

\newcommand{\CS}{coordination sphere}

\newcommand{\paper}{paper~}

\newcommand{\sublattice}[1]{\tilde{#1}}

\newcommand{\XY}{xy}
\newcommand{\XtwoMinusYtwo}{x^2-y^2}

\newcommand{\XZ}{xz}
\newcommand{\YZ}{yz}

\newcommand{\ThreeZTwominusRTwo}{3z^2-r^2}

\newcommand{\degree}{^{\circ}}

\newcommand{\deltaE}{\Delta {\cal{E}}}

\newcommand{\deltaOnSite}{\delta_{on\text{-}site}}

\title{Magnetocrystalline anisotropy in metallic systems: \\
fast and stable estimation in Green's functions formalism}


\author{Ilya~V.~Kashin} \email{i.v.kashin@urfu.ru}
    
\author{Sergei~N.~Andreev}

\affiliation{
\BlankLineAboveAndBelow{
Theoretical Physics and Applied Mathematics Department, Ural Federal University, Mira Str. 19, 620002 Ekaterinburg, Russia
}
}

\date{\today}

\begin{abstract}
In this work we suggest a theoretical approach, that allows to study the effects of magnetocrystalline anisotropy (MCA) in metallic systems using the Green's functions formalism. 
We demonstrate that employment of the reciprocal space resolution instead of its reduction in the inter-site variant essentially improves the numerical stability of MCA energy by means of Monkhorst-Pack grid density and spatial convergence. 
The latter problem is able to be completely removed due to rigorous analytical replacement of pairwise atomic summation by simple composition of sublattices contributions, calculated as a whole. 
The approach is validated on the effective model of single atom, which nevertheless inherits the qualitative MCA picture of Co monolayer and Au/Co/Au sandwiched material.
The numerical convergence is confirmed using the model of atomic chain in the strong metallic regime.
For cobalt monoxide, described by \textit{ab initio} calculations using GGA\textit{+U}, the MCA energy angular profile reveals the prevailing role of ferromagnetically aligned Co sublattices in forming of the easy axis.
\end{abstract}



\maketitle



\section{Introduction}

Today, an increasing amount of scientists attention is devoted to magnetic materials exhibiting a non-trivial and stable spin ordering patterns at relatively low temperatures. There is a great number of studies, that report the high sensitivity of their different representatives to the common experimentally available influences, such as external electric or magnetic field and pressure 
\cite{PhysRevB.86.115134, PhysRevB.79.155123, doi:10.1126/science.aau0968, PhysRevApplied.18.044024, Gao2020, Peng2020, Bogdanov2020}.
This remarkable palette of traits opens avenues for designing energy-efficient and high-performance computer memory modules based on the spin degrees of freedom 
\cite{doi:10.1126/science.1240573, doi:10.1126/science.1214143, doi:10.1126/sciadv.aat0415}.

The emergence of the topologically protected magnetic excitations are primarily reasoned by violation of spatial symmetry in the crystal lattice. It follows from the fact that commonly lattice appears as a dominant source of spatial heterogeneity of the electron subsystem of the material. On the contrary, if the symmetry is high, the magnetic picture usually could be exhaustively captured on the level of isotropic pairwise spin-spin interactions of the Heisenberg type. 

In this context, spin-orbit coupling plays a crucial role, inducing a complex interplay between the spin with the distorted crystal field. The latter can be induced by either the mechanical deformation of the crystal 
\cite{PhysRevB.79.155123, PhysRevB.86.115134, DROVOSEKOV2017305, PhysRevB.98.144411} 
or the qualitative reduction of its dimensionality as layered and two-dimensional structures are synthesized 
\cite{Kashin_2020, PhysRevB.108.L161402, ESTERAS2023100072}. 
The indicative outcome is the appearance of energetically favorable directions for atomic moments – easy axes and magnetization planes.

Thus, the energy of magnetocrystalline anisotropy (MCA) becomes an important characteristic for analyzing the possibility of magnetic ordering and its type 
\cite{Kashin_2020, pub.1121024165, pub.1121304002, PhysRevLett.96.147201, PhysRevB.81.054433, PhysRevB.80.195420}. 
For two-dimensional materials, this factor often appears decisive in overcoming the constraints of the Mermin-Wagner theorem 
\cite{PhysRevLett.17.1133} 
regarding the impossibility of stabilizing magnetic order solely through isotropic magnetic interactions.

Once the spin-orbit coupling is weak, this energy can be estimated in the formalism of local magnetic moments of atoms  
as the difference in the total energies of the crystal, found for two cases: atomic magnetic moments aligned with the quantization axis and atomic magnetic moments oriented at an arbitrary angle to this axis. 
This approach has proven effective in studies based on first-principles calculations
\cite{PhysRevB.86.115134, Alsaad_2018, PhysRevMaterials.6.024402, PhysRevB.78.054413} 
but lacks flexibility in addressing the structural elements of the crystal.

It comes as a reason why there is yet another popular approach, which is based on Green's functions framework. Starting from the crystal's Hamiltonian, written in the tight binding approximation, with the spin-orbit coupling treated as the perturbation, an expression for the MCA energy could be derived. It has contributions of all crystal's atoms in the form of pairwise terms
\cite{PhysRevB.52.13419, PhysRevB.89.214422, Grytsiuk2020}. 
However, the practical application of such an approach is generally limited to insulating systems, for which it is sufficient to consider the spatial surrounding up to the third nearest neighbors for each atom. 
Conducting systems lose this advantage, making the numerical convergence of the pairwise contributions sum troublesome and often non-achievable, as well as it was stated in the case of building the picture of Heisenberg exchange interactions environment
\cite{PhysRevB.64.174402, PhysRevB.106.134434, Kashin_JMMM}.
Furthermore, it is known to have a stability issues by means of Monkhorst-Pack~\cite{Monkhorst_Pack} grid used to perform the numerical estimations~\cite{PhysRevB.52.13419}. 
Also noteworthy, that examination the MCA energy for the minimum regarding the spacial direction $(\PhiTetta)$ requires corresponding grid search, where obtaining a single $(\PhiTetta)$ point could take hours of multiprocessing computation. In this view, for the investigation of MCA in real materials the fast algorithm comes as strictly necessary.

Due to these challenges, this work is aimed to develop a method based on the application of spatially-dependent Green's functions to estimate the MCA energy for arbitrary direction $(\PhiTetta)$. For the purpose of easy axis (or plane) determination we map the whole angular profile onto the model of non-interacting spins, where MCA is described by single ion anisotropy tensor (SIAT). We demonstrate that it not just enhances numerical stability, but also reduces the calculation time by two orders of magnitude on a modern computer.

First of all, the suggested approach is tested on the simplest model of a single atom, for which the character of the MCA can be predicted from the general physical principles. 
After that on the case of an one-dimensional atomic chain in the strong metallic phase we demonstrate the stability of this approach with respect to the $\bm{k}$-grid density. The latter model appeared illustrative to show that thus found numerical values of the MCA energy could be stated as the asymptotic ones for the expression based on the spatial summation of the pairwise atomic contributions.

Approbation of the approach on the real materials is performed on the case of transition metal monoxide CoO. In the paramagnetic phase, this compound crystallizes in rock-salt structure. However, the formation of antiferromagnetic long-range ordering, which is classified as AFM II~\cite{PhysRev.110.1333}, at low temperatures causes significant distortions, which lowers the symmetry of the crystal field to a monoclinic structure. This monoclinic distortion could be described as a composition of a
rhombohedral and an orthorhombic one, and together with the partially filled $t_{2g}$ orbitals favours CoO to generate strong MCA of the easy axis character.
However, experimental and theoretical studies~\cite{PhysRevB.64.052102, PhysRevB.86.115134} give ambiguous estimations of the spacial direction of this axis, and the mechanisms of its formations are yet to be understood.

In this work CoO is studied on the base of first-principles calculations, performed in generalized-gradient approximation with intra-atomic Coulomb interaction taken into account (GGA\textit{+U}). To describe the MCA we perform the decomposition of MCA energy angular profile, obtained for one Co atom, onto individual contributions of each sublattice.
The sublattice should be understood as the composition of Co atoms with the same local positions in the unit cells (each sublattice geometrically is the Bravais lattice with atoms in the sites).

It reveals that easy axis of this atom (on the background of the on-site effects) is only reasoned by Co-Co atomic interplay in (101) plane, where their spin magnetic moments are collinear. Therefore, we expect the technologically perspective behaviour of CoO MCA if an additional mechanism of lattice distortion along (101) is introduced, such as a directional pressure.



\section{Method}


\subsection{Finding SIAT}


Single ion anisotropy tensor is defined in frame of the spin model, where each magnetic atom (indexed as~$i$) is presented as the corresponding classic vector 
$\bm{S}_{i}~=~\big| \bm{S}_{i} \big| \, 
(\sin{\theta_{i}} \cos{\varphi_{i}}, \,
 \sin{\theta_{i}} \sin{\varphi_{i}}, \,
 \cos{\theta_{i}} )$.
The Hamiltonian of such model could be written as a simple combination of expressions, which set some spatial axis or plane as favorable for each spin individually. For our purpose we take $\theta_{i} = \theta$, $\varphi_{i} = \varphi$ and thus establish the model as the energy of all spins, simultaneously canted on $(\PhiTetta)$:
%
\begin{equation}
\label{Eq:Spin_Hamiltonian}
    \SpinHamiltonianSIAT (\PhiTetta) = 
        \sum_{i} \,  \big| \bm{S}_{i} \big|^{2} \,
        \sum_{\mu \nu} \,
            e_{i}^{\mu} \,
            A_{i}^{\mu \nu} 
                e_{i}^{\nu}
                \eqcomma
\end{equation}
where 
$e_{i}^{\mu}~=~S_{i}^{\mu} / \big| \bm{S}_{i} \big|$
and
$A_{i}^{\mu \nu}$ is the SIAT element ($\mu, \nu = x, y, z$).

Then to describe the crystal in electron language the common practise is to use the tight-binding approximation~\cite{Goringe_1997}. Assuming that spin-orbital coupling is neglected, we write:
\begin{equation}
\label{Eq:TBModel}
\begin{split}
      \hat{H} =
            \sum_{i \ne j}
                  \sum_{\alpha \beta}
                      & \sum_{\sigma}
                          \, t^{\sigma}_{i(\alpha) \, j(\beta)}
                                    \,
                              \hat{c}^{\dagger}_{i(\alpha) \sigma}
                                        \hat{c}_{j(\beta)  \sigma}
             + \\
            &+
            \sum_{i}
                  \sum_{\alpha}
                        \sum_{\sigma}
                           \, \varepsilon^{\sigma}_{i(\alpha)}
                                    \,
                              \hat{c}^{\dagger}_{i(\alpha) \sigma}
                                        \hat{c}_{i(\alpha) \sigma} \eqcomma
\end{split}
\end{equation}
where 
$\hat{c}^{\dagger}_{i(\alpha) \sigma}$ is creation operator of the electron with the spin $\sigma$ on the orbital $\alpha$ of the atom $i$; 
$\hat{c}_{j(\beta) \sigma}$ is
annihilation operator of the electron with the spin $\sigma$ on the orbital $\beta$ of the atom $j$;
$t^{\sigma}_{i(\alpha) \, j(\beta)}$ is the electron hopping integral and 
$\varepsilon^{\sigma}_{i(\alpha)}$ is the on-site energy of the electron.

This Hamiltonian could be equally represented in matrix form with illustrative and fruitful interpretation. If the on-site physics and hybridization picture is being explained on the level of interplay between two different unit cells, one can then state one of this cell as having zero translation vector $\bm{T} = 0$ (i.e. choose the coordinate system origin) and express how it couples with some another cell with translation $\bm{T}$:
\begin{equation}
    \big[
    H^{\sigma}(\bm{T})
    \big]_{ij} 
    = 
    t_{ij}^{\sigma}
    + 
    \varepsilon_{i}^{\sigma} \, 
    \delta_{ij} \, 
    \delta_{\bm{T} 0} \eqdot
\end{equation}
where
$H^{\sigma}(\bm{T})$ is the unit cell sized matrix, whereas indices $ij$ denote selection of $i$ atom from $\bm{T} = 0$ cell and $j$ atom from $\bm{T}$ cell, interplay of which appear as the sector of this matrix. Naturally, the case of the both cells being actually the same one with $\bm{T} = 0$ includes, apart from intra-cell atomic couplings, the on-site electron energies. 

This matrix form is convenient to be presented in the reciprocal space
\begin{equation}
      H^{\sigma}(\bm{k}) = \sum_{\bm{T}} \, H^{\sigma}(\bm{T}) \cdot \regexp(i \bm{k} \bm{T})
\end{equation}
to introduce the Green's functions toolbox.
%
%
%
%
According to the definition, we write the expression for $\bm{k}$-dependent Green's function:
\begin{equation}
      \GkE = \big\{ \energy - H^{\sigma}(\bm{k}) \big\} ^{-1} \eqcomma
\end{equation}
where 
$\energy$ denotes the spectrum sweep energy as the diagonal matrix of Hamiltonian matrix size. Then the inter-site version, describing the coupling between atoms $i$ and $j$, reads
\begin{equation}
\label{Eq:Gij}
    G^{\sigma}_{ij} = 
    \frac{1}{N_{\bm{k}}} 
    \sum_{\bm{k}} 
    \big[ \GkE \big]_{\sublattice{i} \sublattice{j}}
    \cdot 
    \regexp(-i \bm{k} \bm{T}_{ij})  \eqcomma
\end{equation}
where 
$\bm{T}_{ij}$ is the translation vector, which connects the unit cells of these atoms,
$N_{\bm{k}}$ is the number of Monkhorst-Pack grid points~\cite{Monkhorst_Pack}
and $\big[ \GkE \big]_{\sublattice{i} \sublattice{j}}$ is the sector of unit cell sized matrix, which could be interpreted as the interplay between two sublattices of the crystal. 

The concept of sublattice goes as following. Each atom in the unit cell has its own local position in this cell. And the group of atoms from different cells but with the same local positions compose the sublattice, where all distances are fully defined by translation vectors $\bm{T}$. Therefore, the crystal itself could be presented as the entirety of sublattices, amount of which is simply the number of atoms in the unit cell. For the further analysis it is crucially to note that 
$\big[ \GkE \big]_{\sublattice{i} \sublattice{j}}$ alone actually depends on the \textit{local} positions of atoms $i$ and $j$ in their unit cells (the tilde denotes corresponding sublattice), but not on $\bm{T}_{ij}$. 

If under our consideration are only $3d$ systems, the spin-orbit coupling (SOC) could be added to initial electron model, \EqRef{Eq:TBModel}, as a perturbation 
\cite{PhysRevB.39.865, PhysRevB.52.13419}.
In this framework MCA energy along 
quantization axis (commonly denoted as $z$) reads 
\cite{PhysRevB.89.214422}
\begin{equation}
\label{Eq:ElectronMAE}
\begin{split}
    \ElectronMAE &= \\
     - \frac{1}{2 \pi} 
  &    \sum_{ij}
       \IntImTrLSigma
            {
            H^{so} \, G_{ij} \, H^{so} \, G_{ji}
            }
       \eqcomma
\end{split}
\end{equation}
where
$\energy_{F}$ is the Fermi energy,
$\mathrm{Tr_{L \, \sigma}}$ is the trace over orbital ($\mathrm{L}$) and spin ($\mathrm{\sigma}$) indices,
the Green's functions are expressed in a spinor form 
$G_{ij} =
\bigl(
\begin{smallmatrix}
    G_{ij}^{\uparrow}   &          0             \\
          0             &    G_{ij}^{\downarrow}
\end{smallmatrix}
\bigr)
$
and the SOC operator for $d$ shell
$H^{so}~=~
\lambda \, 
\bm{{\cal{L}}} 
\bm{{\cal{S}}}~=~
\bigl(
\begin{smallmatrix}
    [H^{so}]^{\uparrow \uparrow}     &    [H^{so}]^{\uparrow \downarrow}      \\
    [H^{so}]^{\downarrow \uparrow}   &    [H^{so}]^{\downarrow \downarrow}
\end{smallmatrix}
\bigr)
$
remains constant and equal for any atom taken into account (with $\lambda$ as the small parameter).

Finding MCA energy along some another direction
$(\sin{\theta} \cos{\varphi}, \,
  \sin{\theta} \sin{\varphi}, \,
  \cos{\theta} )$
requires corresponding rotation from $(0, 0, 1)$, applied to $H^{so}$:
\begin{equation}
\label{Eq:HsoRotation}
    H^{so}(\PhiTetta) = U^{-1}(\PhiTetta) \, H^{so} \, U(\PhiTetta)   \eqcomma
\end{equation}
where
\begin{equation}
    U(\PhiTetta) = 
    \begin{pmatrix}
    \phantom{-} \regcos{(\theta / 2)} \phantom{ \cdot e^{i \varphi} } &
                \regsin{(\theta / 2)} \cdot e^{-i \varphi}              \\ 
      - \regsin{(\theta / 2)} \cdot e^{i \varphi} & 
        \regcos{(\theta / 2)} \phantom{ \cdot e^{-i \varphi} }
    \end{pmatrix}
\end{equation}
is Wigner's rotation matrix. Thus established $\ElectronMAE(\PhiTetta)$ could be fitted by \EqRef{Eq:Spin_Hamiltonian} in terms of $A_{i}^{\mu \nu}$ for particular set of $\big| \bm{S}_{i} \big|$:
\begin{equation}
\begin{split}
   [&A_{i}^{\mu \nu}]^{*} = \\
    &\underset{ \{ A^{\mu \nu} \} }{\mathrm{argmin}} 
        \Big[
           \iint_{\PhiTetta} \, 
                \big|\big|
                    \SpinHamiltonianSIAT(\PhiTetta) - \ElectronMAE(\PhiTetta)
                \big|\big|
                       \, d\varphi \, d\theta
        \Big]
    \eqdot
\end{split}
\label{eq:SIATcalculation}
\end{equation}






\subsection{Optimization}


First of all, we should note that MCA energy, \EqRef{Eq:ElectronMAE}, in fact represents the simple composition of individual magnetic pictures, centered on different $i$ atom. The explicit indication 
$\ElectronMAE(\PhiTetta) = \sum_{i} \{ \ElectronMAE(\PhiTetta) \}_{i}$ enables one to directly match them with the terms of spin Hamiltonian, \EqRef{Eq:Spin_Hamiltonian}.

Therefore, finding of each $\{ \ElectronMAE(\PhiTetta) \}_{i}$ requires successive consideration of all crystal's atoms (indexed as $j$) in a pairwise manner. In the case of insulating systems, where short-ranged micro-magnetic interactions prevail, it appears enough to take into account the atoms up to the third \CS~to obtain the converged result 
\cite{PhysRev.94.1498, Goringe_1997, PhysRevB.106.134434}.
On the contrary, metallic systems, due to conduction electrons, demonstrate long-ranged behavior and hence strongly decelerate the convergence 
\cite{PhysRevB.91.125133, PhysRevLett.116.217202, PhysRevB.64.174402, Kashin_JMMM, PhysRevB.106.134434}.
Even assuming it to be possibly reached (by no means guaranteed), we should consider hundreds of {\CS s} to obtain $\{ \ElectronMAE(\PhiTetta) \}_{i}$ for one particular set of $\PhiTetta$ and $i$. The situation is additionally worsened by the fact, that numerical stability of \EqRef{Eq:ElectronMAE} appears sensible to the density of $\bm{k}$-points grid the Green's functions are calculated on 
\cite{PhysRevB.52.13419}. 
It dramatically escalates the expected calculation time on the modern multiprocessor computer and rises the technical demands to the required RAM.

To find the comprehensive solution in this \paper we suggest to rewrite the expression for MCA energy, \EqRef{Eq:ElectronMAE}, in terms of local $\bm{k}$-dependent Green's functions instead of inter-site ones. As the first step for the sake of brevity let us define an operator:
\begin{equation}
    \Operator[\bullet] = - \frac{1}{2 \pi} \, \IntImTrLSigma{\bullet} \eqdot
\end{equation}
Then derivation starts from \EqRef{Eq:ElectronMAE} with explicitly highlighted descriptions of $\{ \ElectronMAE(\PhiTetta) \}_{i}$:
\begin{equation}
\begin{split}
    \ElectronMAE & (\PhiTetta) = \\
       &\sum_{i} \,\,
            \Operator
                \Big[
                    \sum_{j}  H^{so}(\PhiTetta) \, G_{ij} \, H^{so}(\PhiTetta) \, G_{ji}
                \Big]
        \eqdot
\end{split}
\end{equation}
Let us substitute the expanded form of $G_{ij}$ and $G_{ji}$ from \EqRef{Eq:Gij} into the argument of $\Operator$. Here we employ the independence of $\big[ \GkE \big]_{\sublattice{i} \sublattice{j}}$ from $\bm{T}_{ij}$:
\begin{equation}
\begin{split}
  \sum_{j}  H^{so}(\PhiTetta) \, & G_{ij} \, H^{so}(\PhiTetta) \, G_{ji} = \\
  \frac{1}{ N_{\bm{k}} } \frac{1}{ N_{\bm{k'}} }
     \sum_{ \sublattice{j} } &  \sum_{ \bm{k} \bm{k'} }    \\
  \times \, H^{so}(\PhiTetta) & \, \big[ \GkEspinor      \big]_{\sublattice{i} \sublattice{j}} \,
            H^{so}(\PhiTetta)   \, \big[ \GkhatchEspinor \big]_{\sublattice{j} \sublattice{i}} \, \\
  \times \, \sum_{ \bm{T}_{ij} } \, 
    \regexp &
    \big(
         i \, \{ \bm{k}' - \bm{k} \} \, {\bm{T}}_{ij}
    \big)
        \eqdot
\end{split}
\end{equation}
The last sum could be found analytically:
\begin{equation}
    \sum_{ \bm{T}_{ij} } \, 
        \regexp  
        \big(
             i \, \{ \bm{k}' - \bm{k} \} \, {\bm{T}}_{ij}
        \big)
         = N_{\bm{k}} \cdot \delta(\bm{k}' - \bm{k})
\end{equation}
and follows us to the final expression:
\begin{equation}
\label{Eq:FinalMAE}
\begin{split}
    \ElectronMAE (\PhiTetta) = \phantom{space} &  \\
        \sum_{i} \,\, \frac{1}{ N_{\bm{k}} }  \,
            \Operator
                \Big[
                    \sum_{ \sublattice{j} }
                    \sum_{ \bm{k} } \,
                      & H^{so}(\PhiTetta)  \, 
                        \big[ \GkEspinor \big]_{\sublattice{i} \sublattice{j}} \, \\
                 \times & H^{so}(\PhiTetta)  \, 
                        \big[ \GkEspinor \big]_{\sublattice{j} \sublattice{i}} \,
                \Big]
        \eqdot
\end{split}
\end{equation}
The prime advantage of this result is emphasized to be vanishing of the direct spacial sum over all crystal's atoms $j$ with the subsequent replacement by the composition of sublattice-originated terms $\sublattice{j}$ in a little amount of a number of atoms in the unit cell. Therefore, in view of the fact that for metallic systems the former sum contains hundreds of non-negligible contributions, the numerical calculation of $\ElectronMAE$ on some grid of $(\PhiTetta)$ using \EqRef{Eq:FinalMAE} is expected to be two orders of magnitude faster than the same employment of \EqRef{Eq:ElectronMAE} with \EqRef{Eq:HsoRotation}.

Furthermore, \EqRef{Eq:FinalMAE} possesses robuster stability by means of $\bm{k}$-grid density. In order to show it let us carry out the formal analysis in a following way. Any considered inter-site Green's function, calculated on some practically implementable grid with $N_{\bm{k}}$ $\bm{k}$-points, we can express as the deviation from the "ideal" case $N_{\bm{k}} = \infty$. Naturally, if the "real" grid is dense enough, this deviation should be stated small. One thus can write
\begin{equation}
G_{ij}(N_{\bm{k}}) = G_{ij}(\infty) + \delta G_{ij}(N_{\bm{k}})
\eqdot
\end{equation}
Substituting it into \EqRef{Eq:ElectronMAE} results in
\begin{equation}
\begin{split}
    \ElectronMAE = 
        \sum_{ij} \, 
      & \Operator 
        \big[ 
            H^{so} \, G_{ij}(\infty) \, H^{so} \, G_{ji}(\infty)
        \big] \\
 + \, & \Operator 
        \big[ 
            H^{so} \, \delta G_{ij}(N_{\bm{k}}) \, H^{so} \, \delta G_{ji}(N_{\bm{k}})
        \big] \\
 + \, & \Operator 
        \big[ 
            H^{so} \, G_{ij}(\infty) \, H^{so} \, \delta G_{ji}(N_{\bm{k}})
        \big] \\
 + \, & \Operator 
        \big[ 
            H^{so} \, \delta G_{ij}(N_{\bm{k}}) \, H^{so} \, G_{ji}(\infty)
        \big]
            \eqdot
\end{split}
\end{equation}
Here the first term corresponds to theoretically exact $\ElectronMAE$ and the second term could be assumed negligible as proportional to the square of small deviation. Whereas the third and the fourth term appear linearly dependent from this deviation, notably amplified by $G_{ij}(\infty)$ and $G_{ji}(\infty)$, correspondingly. This fact stands as a clear explanation why for non-insulating real materials the $\ElectronMAE$ might not approach the convergence by means of technically available $\bm{k}$-grid density 
\cite{PhysRevB.52.13419}. 
In comparison with \EqRef{Eq:ElectronMAE}, our final expression, \EqRef{Eq:FinalMAE}, contains only one sum over $\bm{k}$-points, which eliminates the possibility to escalate the finite $\bm{k}$-grid reasoned deviation by Green's functions themselves. 
In the next section we show that it indeed results in more robust convergence, even for metallic systems.



\section{Implementation}


\subsection{Single atom model}


The first and most important step for validation of the suggested MCA energy expression, \EqRef{Eq:FinalMAE}, should be representation of the simplest possible model for which the anisotropy direction can be predicted on the basis of general physical considerations. The problem of constructing such a model appears non-trivial, since it is necessary to design the spatial inhomogeneity of the orbital moment and perform its transmission to the spin moment.

Obviously, we should start with the remark that observation of MCA requires uncompensated spin moment. Consequently, only partially occupied electron orbitals are expected to contribute. Therefore, the simplest band structure (of $3d$ systems) to highlight the MCA could contain five $d$-orbitals, that have one spin channel completely empty or fully occupied, and another one - nearly half-filled. 

In addition, the bandwidth is known to play an important role. Traditionally, it is understood as proportional to the net hybridization of an orbital with its surroundings. But to satisfy our goal of simplicity, we can confine its representation within only one $\Gamma$ point in the reciprocal space. In this frame the bandwidth could be approximated by on-site energy difference between $d$-orbitals. It grants a remarkable possibility to completely remove the crystal field from focus, turning the crystal model into a model of the single atom with five free parameters only - the on-site electron energy on $d$-orbitals with the spin projection, which corresponds to the nearly half-filling case (another spin channel is tuned to be fully occupied also by means of corresponding on-site electron energies).

Thus, one can clearly distinguish the orbitals $\XY$ and $\XtwoMinusYtwo$ as lying in the plane perpendicular to the quantization axis $z$ (in-plane orbitals),
and the orbitals $\XZ$, $\YZ$ and $\ThreeZTwominusRTwo$ lying outside this plane (out-of-plane orbitals). 

Assuming that through the spin-orbit interaction mechanism in-plane and out-of-plane orbitals hybridize independently of each other, the number of free parameters can be reduced to two: in-plane and out-of-plane splitting. In order to provide the half-filling occupancy for one spin channel we define the energies as follows 
\begin{equation}
\begin{split}
   &\varepsilon^{\uparrow}(\XtwoMinusYtwo) = -V_{\parallel}                \\
   &\varepsilon^{\uparrow}(\XY) = +V_{\parallel}                           \\
   &\varepsilon^{\uparrow}(\ThreeZTwominusRTwo) = -V_{\perp} \; ,  \\
   &\varepsilon^{\uparrow}(\XZ) = +V_{\perp}                       \\
   &\varepsilon^{\uparrow}(\YZ) = +V_{\perp}                 
\end{split}
\label{eq:SingleAtomModel}
\end{equation}
where $2 V_{\parallel}$ and $2 V_{\perp}$ are the in-plane and out-of-plane splittings, correspondingly, spin "down" is assumed fully occupied and the Fermi level is zero. 

It is important to note that this model, despite its simplicity, turns out to have an fruitful implementations to explain MCA in real materials. For instance, in work \cite{STOHR1999470} authors use the same principles of band structure schematization to qualitatively describe the appearance of in-plane and out-of-plane magnetic anisotropy in a Co monolayer sandwiched between Au layers. The density of states of Co atoms' $d$-shell, found from the first-principles calculations \cite{PhysRevB.50.9989}, indeed shows full occupancy for one spin channel and partial occupancy for the other one.

Authors demonstrate that the direction of MCA is significantly affected by interaction between the layers. In its absence (the case of a free-standing Co layer) the tetragonal symmetry of the crystal field leads the out-of-plane component of the orbital moment to be quenched. Therefore, the orientation of the spin momentum is also expected to be in-plane. The introduction of Au layers from above and below enhances out-of-plane hybridization, which makes corresponding component of the orbital moment predominant and consequently changes the character of magnetic spin anisotropy. 

By relating the width of electron bands to the intensity of in-plane and out-of-plane splitting, authors numerically analyze the MCA character as a function of $V_{\perp} / V_{\parallel}$. The calculations reproduce an in-plane anisotropy at $V_{\perp} / V_{\parallel} < 1$ and an out-of-plane direction in the opposite case.
It successfully reproduces the expectations for real materials, with the given parameters, estimated in the work~\cite{STOHR1999470}:
$V_{\perp}$~=~0.5~eV, $V_{\parallel}$~=~1~eV for the free Co layer 
 and 
$V_{\perp}$~=~1.5~eV, $V_{\parallel}$~=~1~eV for the Au/Co/Au layered structure.

\begin{figure}[htbp]
\includegraphics[width=0.35\textwidth]{\ContentPath{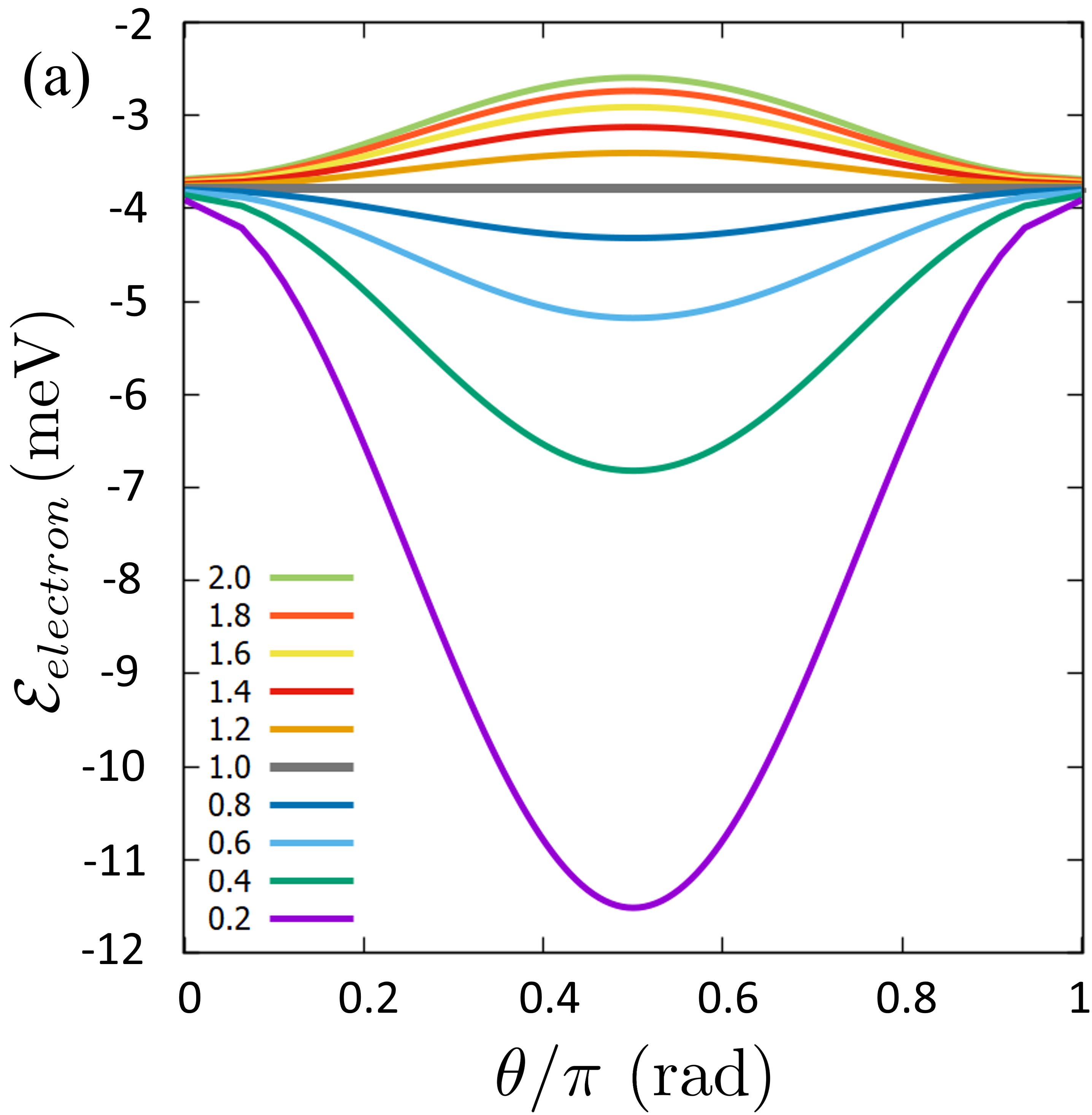}}
\includegraphics[width=0.35\textwidth]{\ContentPath{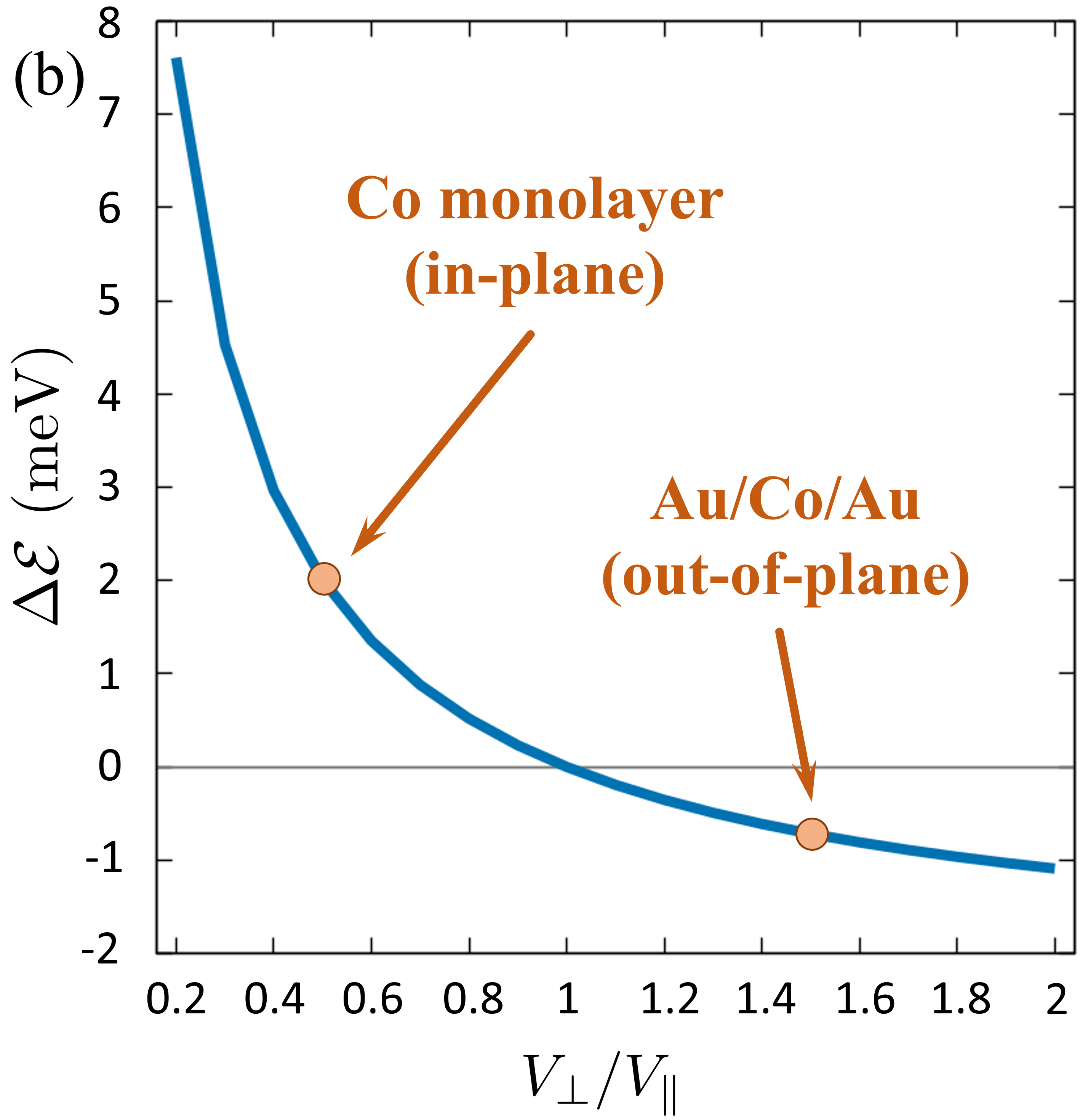}}
\includegraphics[width=0.35\textwidth]{\ContentPath{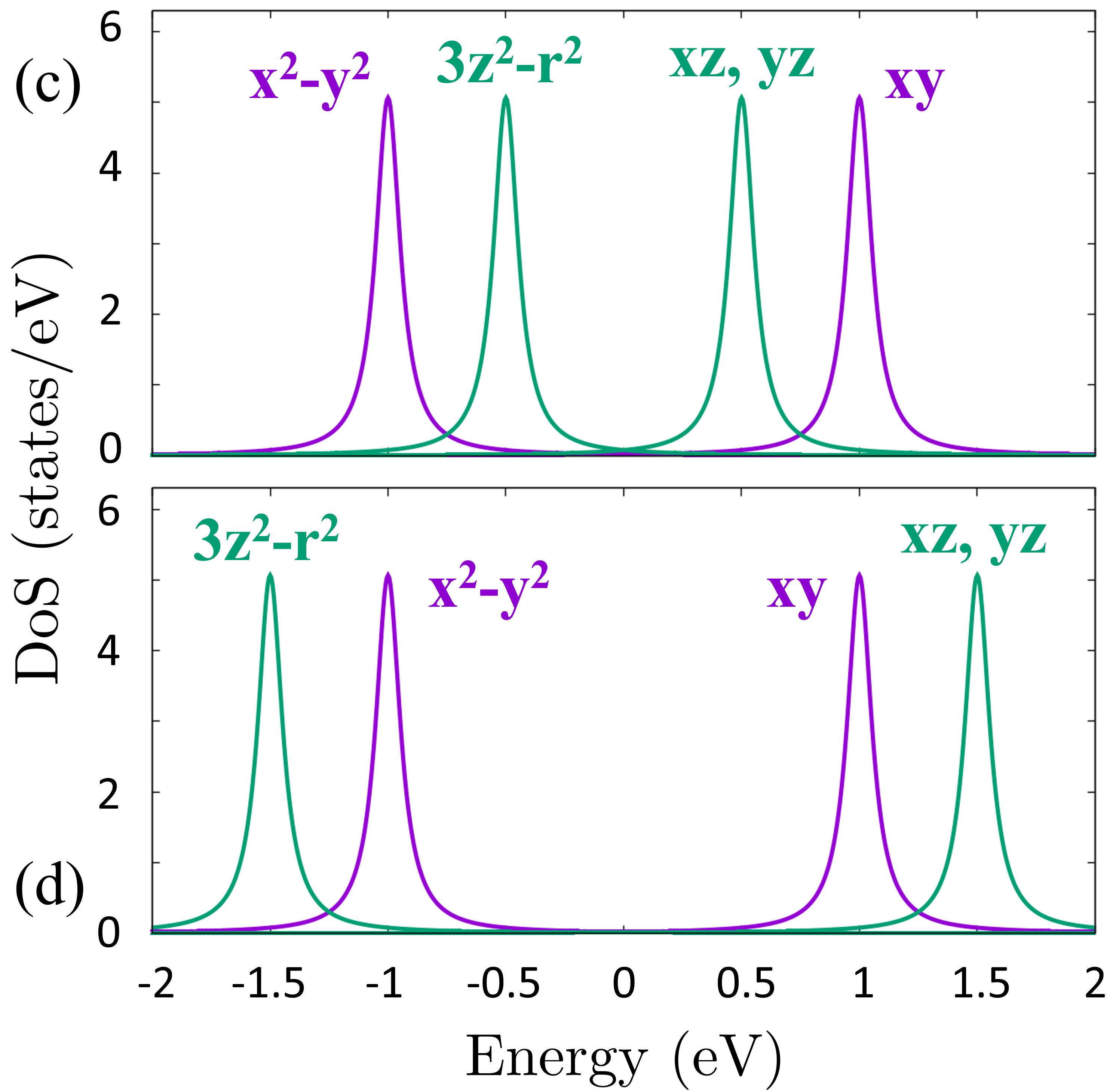}}

\caption{\label{fig:SingleAtom} Single atom model.
\textit{(a)} 
MCA energy as a function of $\theta$, obtained using \EqRef{Eq:FinalMAE} (this energy is degenerated with respect to $\varphi$). In the legend there are correspoding values of $V_{\perp} / V_{\parallel}$.
\textit{(b)} 
MCA energy difference between out-of-plane and in-plane direction, \EqRef{Eq:DeltaE}.
\textit{(c, d)} Density of states, calculated 
for Co monolayer 
($V_{\parallel}$~=~1~eV, $V_{\perp}$~=~0.5~eV) and 
for the Au/Co/Au layered structure 
($V_{\parallel}$~=~1~eV, $V_{\perp}$~=~1.5~eV) (\EqRef{eq:SingleAtomModel}, Fermi level is zero). 
Note that this model does not contain any hopping integrals, \EqRef{Eq:TBModel}, causing complete independence from $\bm{k}$-grid density.
}
\end{figure}

Comprehensive elaboration of considered single atom model using suggested MCA energy expression, \EqRef{Eq:FinalMAE}, is summarized in Fig.~\ref{fig:SingleAtom}. 
One can mention on the density of states spectrum, where the width of each band is approximated by a composition of two (in-plane orbitals) and three (out-of-plane orbitals) single-peaked bands, which are connected to each other through the spin-orbital coupling. 
It is clearly seen, that employing $\bm{k}$-dependent Green's functions perfectly reproduces the same picture of MCA, with rigorous switching of the direction while crossing the $V_{\perp} / V_{\parallel} = 1$ point.
Apart from the qualitative consistence, our approach yields
\begin{equation}
\deltaE = \ElectronMAE(\varphi, \theta =  0 \degree) -
          \ElectronMAE(\varphi, \theta = 90 \degree) 
\label{Eq:DeltaE}
\end{equation}
as $2$~meV 
for the free Co layer
and
$-0.7$~meV for the Au/Co/Au sandwich (with spin-orbital constant $\lambda = 0.07$~eV),
exactly as it is reported in~\cite{STOHR1999470}.
Hereinafter omitting $\varphi$ as an argument denotes the absence of the actual dependency for particular case under consideration.
Obtained using \EqRef{eq:SIATcalculation} SIAT has the diagonal form with 
$[ A_{i}^{xx} ]^{*} = [ A_{i}^{yy} ]^{*} = 
\ElectronMAE(\varphi, \theta = 90 \degree) / \big| \bm{S}_{i} \big|^2$ and 
$[ A_{i}^{zz} ]^{*} = 
\ElectronMAE(\varphi, \theta =  0 \degree) / \big| \bm{S}_{i} \big|^2$,
where $\big| \bm{S}_{i} \big| = 5/2$.


\subsection{Simple metal model}


The next step of our study is to compare the numerical stability of \EqRef{Eq:FinalMAE} with the original approach involving an internal sum over lattice atoms, \EqRef{Eq:ElectronMAE}. 
As manifested, the prime motivation for the new expression is to improve the slow or absent convergence of the internal sum if metallic system is considered. 
Let us extend the single atom model by setting back the natural mechanism of bandwidth formation - hybridization with the orbitals of neighboring atoms, described by hopping  integrals, \EqRef{Eq:TBModel}. 
For this purpose, we are to explicitly take into account the crystal lattice, at least in a minimal concept. 

For the sake of preserving the framework of $V_{\parallel}$ and $V_{\perp}$ parameters as bandwidth tuners, here one can introduce an one-dimensional simple crystal chain with the following traits:
one spin channel ("down") is fully occupied with no hoppings;
for another spin channel on-site electron energies $\varepsilon^{\uparrow}$ are set zero and non-zero diagonal hopping matrices 
$t^{\uparrow}_{i(\alpha) \, j(\beta)} = t^{\uparrow}(\alpha) \, \delta_{\alpha \beta}$ 
are only between the first nearest neighbours. 

Thus, once the connection
\begin{equation}
\begin{split}
   &t^{\uparrow}(\XtwoMinusYtwo) = V_{\parallel} / 2            \\
   &t^{\uparrow}(\XY) = V_{\parallel} / 2                       \\
   &t^{\uparrow}(\ThreeZTwominusRTwo) = V_{\perp} / 2 \; ,      \\
   &t^{\uparrow}(\XZ) = V_{\perp} / 2                           \\
   &t^{\uparrow}(\YZ) = V_{\perp} / 2                
\end{split}
\label{eq:OneDmetalHoppings}
\end{equation}
is provided, we get five independent bands
\begin{equation}
 [ H^{\uparrow}(\bm{k}) ]_{\alpha} = 
    2 \, t^{\uparrow}(\alpha) \,\, \regcos( k_x \cdot a ) \, ,
\end{equation}
where $k_x$ represents 1D reciprocal space, $a$ is the lattice constant, and the bandwidth is $4 \, t^{\uparrow}(\alpha)$, in a correspondence with the single atom model.

\begin{figure}[htbp]
\includegraphics[width=0.35\textwidth]{\ContentPath{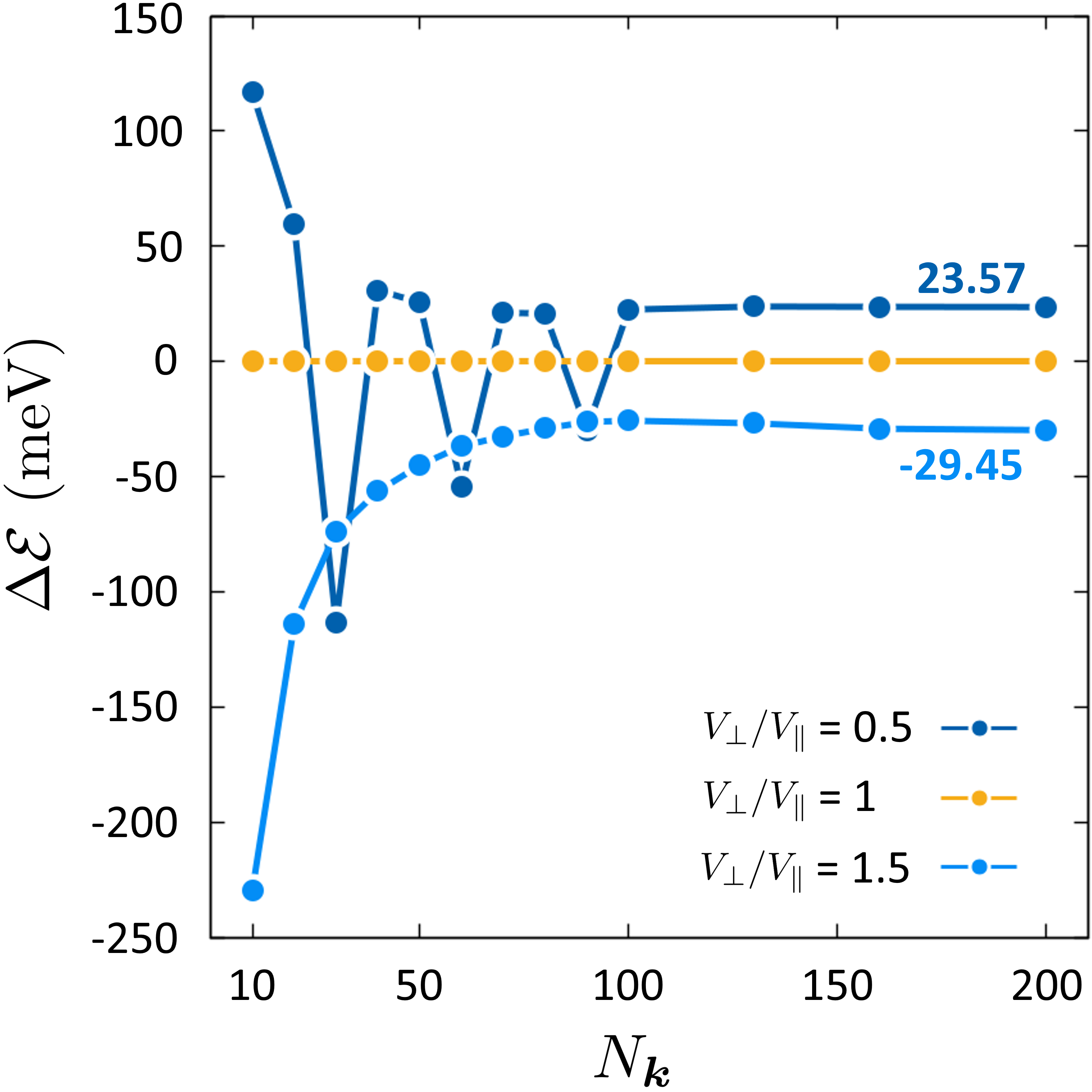}}
\includegraphics[width=0.35\textwidth]{\ContentPath{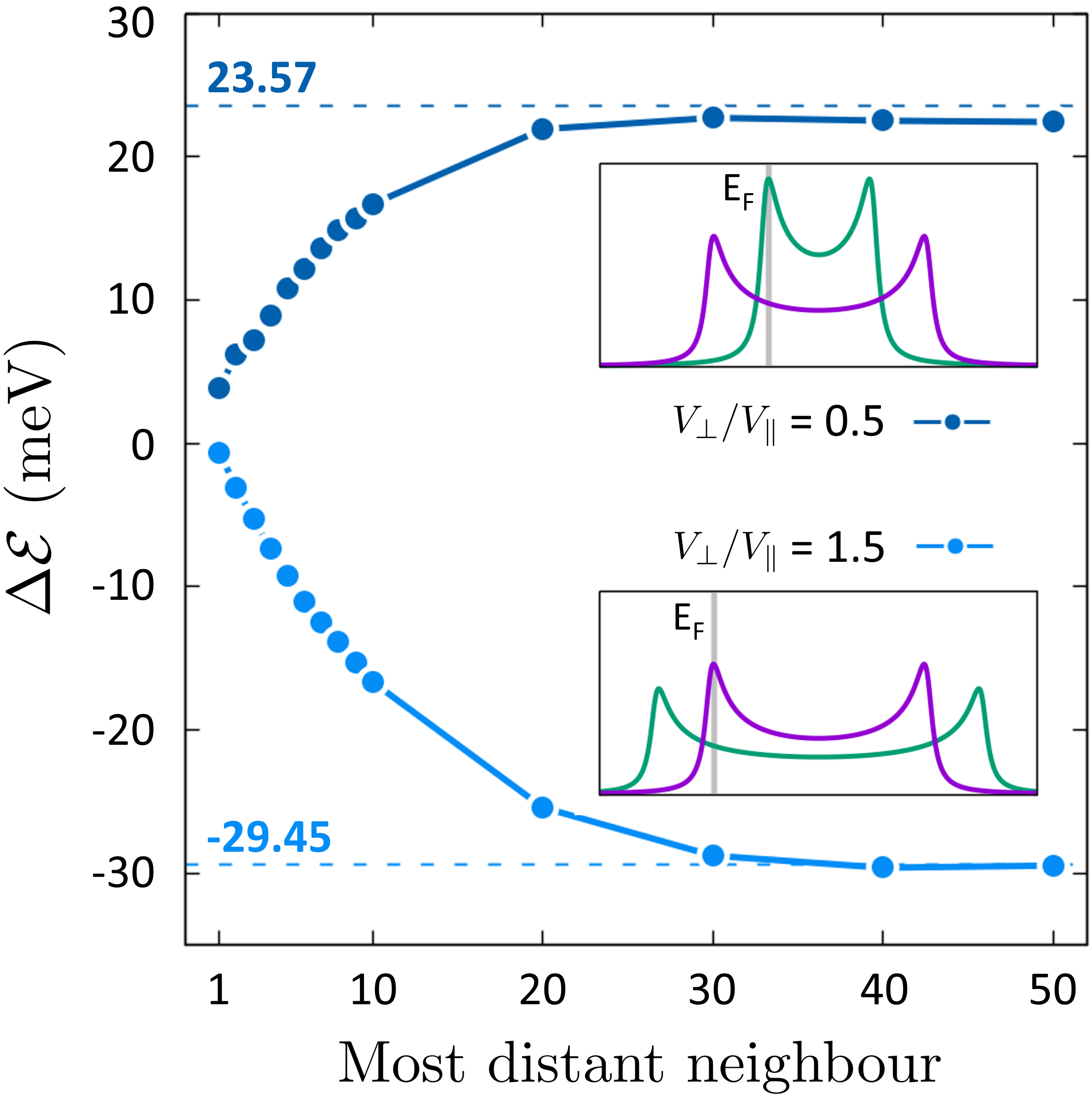}}

\caption{\label{fig:OneDMetal} MCA energy difference between out-of-plane and in-plane direction for the simple 1D metal model, \EqRef{Eq:DeltaE}.
\textit{(top)} 
Dependence from the $\bm{k}$-points amount of the value, obtained using \EqRef{Eq:FinalMAE}.
\textit{(bottom)} 
Employing the \EqRef{Eq:ElectronMAE}, where the pairwise contributions sum is being limited by setting the most distant neighbour taken into account. 
Dashed lines denote the values, obtained using \EqRef{Eq:FinalMAE}.
The $\bm{k}$-points grid is $400~\times~1~\times 1$. 
The \textit{inset} shows the density of states for two considered cases, with the Fermi level adjusted to maximize metallicity. Purple and green lines depict in-plane and out-of-plane $d$-orbitals.
}
\end{figure}

The examination of this model in Green's functions formalism is given in Fig.~\ref{fig:OneDMetal}. 
Here we analyze the same set of $V_{\perp}$ and $V_{\parallel}$ parameters, which corresponds to free Co monolayer and Au/Co/Au sandwich, complemented by 
$V_{\perp} / V_{\parallel} = 1$ case as a critical one.

The first to note is that the two-peaked structure of the density of states curves (Fig.~\ref{fig:OneDMetal}, \textit{bottom, inset}) appears as more physically rigorous representation of the real electronic structure than it was in the single atom model (Fig.~\ref{fig:SingleAtom}, \textit{c-d}), where these bands were approximated by composition of two single-peaked curves. Therefore, we expect the same dynamics of MCA with respect to $V_{\perp} / V_{\parallel}$.

In order to validate the suggested expression stable for metallic systems, for all cases we adjust the Fermi level to be on the peak of density of states, as it is shown in the inset. Fig.~\ref{fig:OneDMetal} (\textit{top}) illustrates that the anticipated tendency of switching the character of MCA in the point $V_{\perp} / V_{\parallel} = 1$ is perfectly reproduced in frame of the simple metal model. One can also observe the reliable convergence of $\deltaE$ as $\bm{k}$-grid density enhances, despite the irregular perturbations on the way. It ensures the applicability of \EqRef{Eq:FinalMAE} for real metallic compounds, which appears especially important on the background of~\cite{PhysRevB.52.13419}, where numerical stability of \EqRef{Eq:ElectronMAE} by means of $\bm{k}$-grid revealed questionable.

Finally, in the Fig.~\ref{fig:OneDMetal} (\textit{bottom}) we study the MCA energy to be estimated using the \EqRef{Eq:ElectronMAE}, as the maximum distance between atoms $i$ and $j$ to contribute to the spacial sum increases. One can see that it takes dozens of coordination spheres to take into account for providing the convergence, whereas the values to approach could be found using suggested expression, \EqRef{Eq:FinalMAE}. Hence, we can again state it valid for the strongly metallic systems, and the basic problem of the spacial sum to be well circumvented.

Finding SIAT reveals that it inherits the same structure as was found for the single atom model: 
$[ A_{i}^{xx} ]^{*} = [ A_{i}^{yy} ]^{*} = 
\ElectronMAE(\varphi, \theta = 90 \degree) / \big| \bm{S}_{i} \big|^2$ and 
$[ A_{i}^{zz} ]^{*} = 
\ElectronMAE(\varphi, \theta =  0 \degree) / \big| \bm{S}_{i} \big|^2$ 
(degeneneracy with respect to $\varphi$ is also preserved). It proves the $\deltaE$ parameter to be decisive to exhaustively capture the MCA picture for the materials, where intra-atomic physics contributes comparably with the spin-lattice interaction.

\subsection{Cobalt monoxide}


It is known that the strong MCA possessed by real bulk crystals is quite rare, due to the high lattice symmetry. The common values of $\deltaE$ have the order of $\mu$eV, but some set of factors met together in one material can boost the value up to a several meV.

A good example is cobalt monoxide CoO, a rock-salt structure of which is distorted by presence of AFM II magnetic order at low temperatures. Accompanied by partially filled $t_{2g}$ orbitals of Co atoms, the robustness of MCA are theoretically showed \cite{PhysRevB.86.115134} to be about 1 meV, generating an easy axis. 

\begin{figure}[htbp]
\includegraphics[width=0.35\textwidth]{\ContentPath{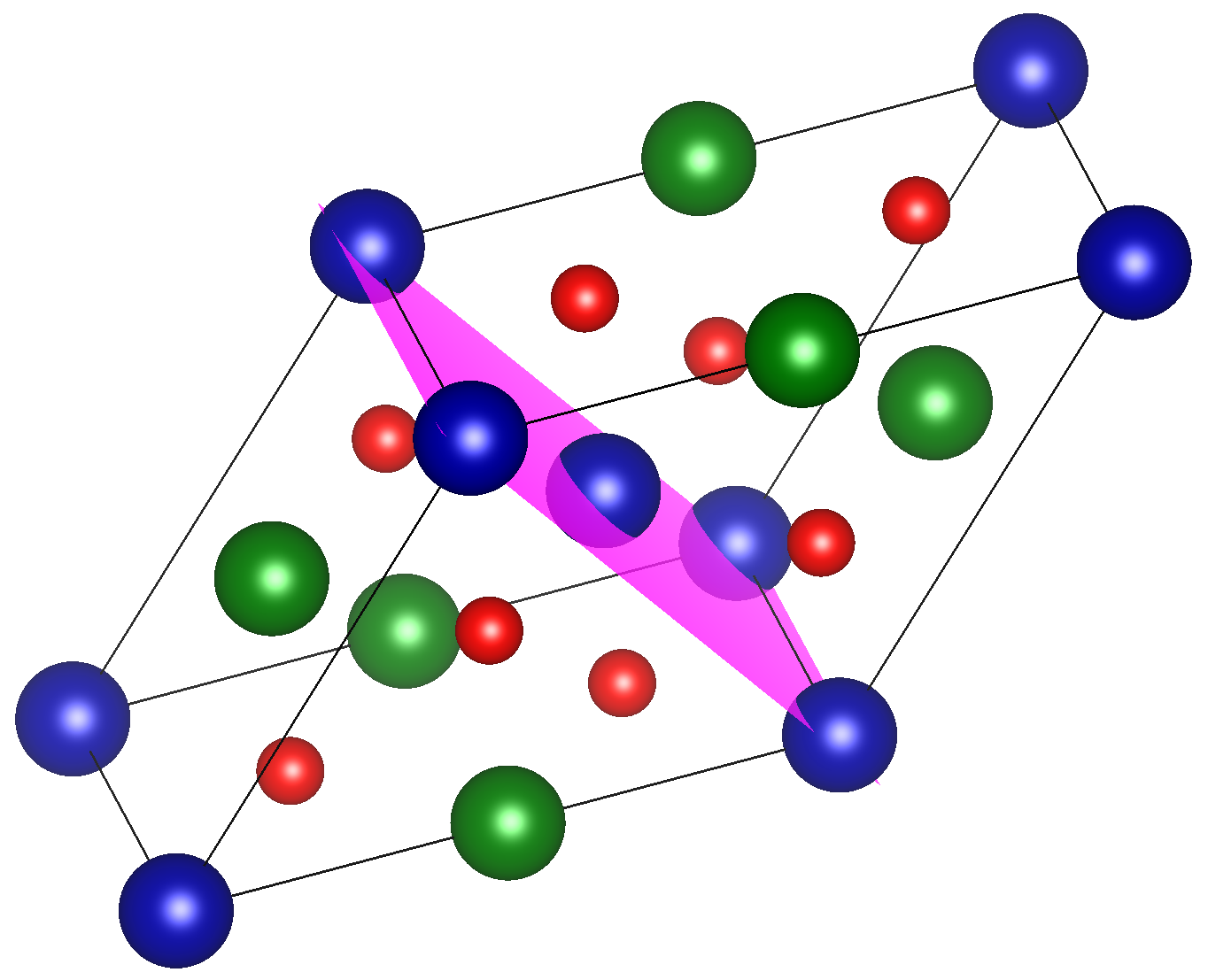}}
\includegraphics[width=0.3\textwidth]{\ContentPath{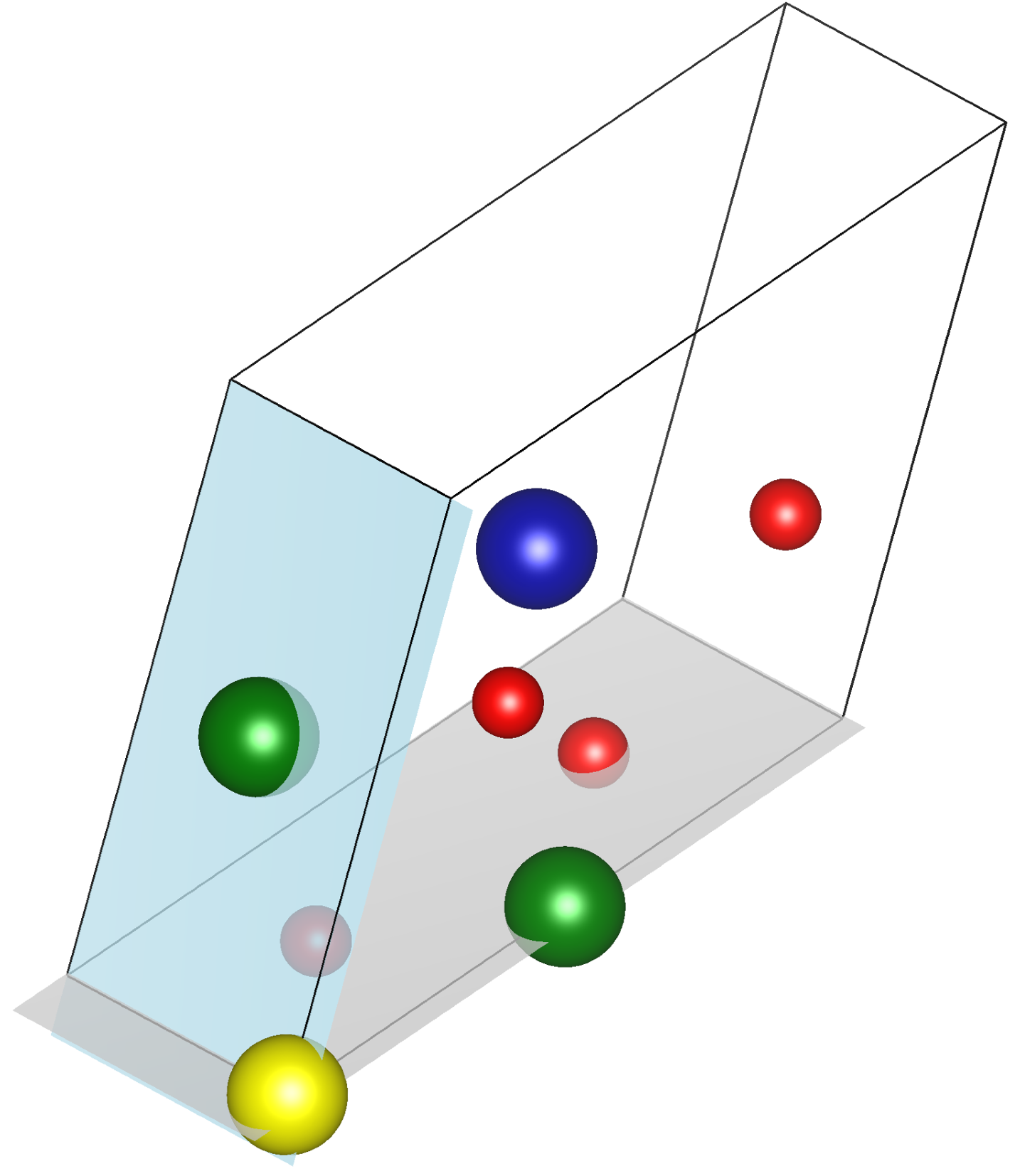}}
\caption{\label{fig:CoOStructure} 
The crystal structure of CoO with monoclinic distortion.
\textit{(top)} Stacking of the ferromagnetic Co planes. The blue (green) spheres represent Co atoms with the spin up (down), and the red spheres denote oxygen atoms. The highlighted plane in our study is (101), thus the stacking is assumed to be along [101].
\textit{(bottom)} Positions of Co atoms in the unit cell. The yellow (\#1) and blue (\#2) spheres are Co atoms with the spin up, and green (\#3 and \#4) spheres show atoms with the spin down. The number in parentheses corresponds to the local positions given in Appendix~\ref{AppendixA}. The yellow color highlights the atom with the local position $(0,~0,~0)$~\AA~and indicates it as for which the MCA energy profile is estimated.
Two planes are depicted to facilitate the visual view.
}
\end{figure}

The monoclinic crystal structure of CoO is illustrated in Fig.~\ref{fig:CoOStructure} (\textit{top}). One can clearly see the stacking of the ferromagnetic planes along [101] with opposite direction of local spin magnetic moments for neighbouring planes, which constitutes the AFM II magnetic order.
Important to mention that the stacking direction appears different from the ordinary cubic [111]~\cite{PhysRev.110.1333}, thus in the text below all corresponding results from references we translate into the coordinate system with the stacking along [101].

The spatial configuration of Co atoms in the unit cell is presented in Fig.~\ref{fig:CoOStructure} (\textit{bottom}). Each of four atoms in the cell thus belongs to different crystal's sublattice, highlighting the fact that the physical lattice of the Co atoms can be understood as the composition of four embedded Bravais lattices.


To elaborate the question of MCA in CoO we perform the first-principles calculations using the GGA\textit{+U} method~\cite{PhysRevLett.77.3865}.
The low-energy model of magnetoactive Co $3d$-shell is obtained in the basis of Wannier functions
~\cite{PhysRevB.56.12847, PhysRevB.65.035109, MOSTOFI2008685}, 
and expressed as its tight-binding Hamiltonian, \EqRef{Eq:TBModel}. Details are provided in the Appendix~\ref{AppendixA}.

Spin magnetic moment per Co atom is obtained as 2.53~$\mu_B$, which is in a good agreement with previous estimations reported
~\cite{PhysRevB.69.155107, PhysRevB.61.5194, PhysRevB.74.155108, PhysRevB.86.115134}.
Nevertheless, various exchange functionals are shown to produce the value closer to 3~$\mu_B$~\cite{PhysRevB.74.155108}. It enables us to treat the on-site spin magnetic moment of each Co atom as spin $\big| \bm{S}_{i} \big| = 3/2$ for parameterize the model, \EqRef{Eq:Spin_Hamiltonian}, by means of SIAT.

The Monkhorst-Pack grid in use is $20 \times 20 \times 20$, which commonly appears as a reliable density to produce physically valid numerical estimations
\cite{PhysRevB.106.134434, Kashin_JMMM} for transition metal bulk magnets.
Nevertheless, for the Co atom with the local position 
$(0,~0,~0)$~\AA~(see Fig.~\ref{fig:CoOStructure} (\textit{bottom}) and Appendix~\ref{AppendixA})
the computation of MCA energy full angular profile on $\varphi \times \theta$ grid with density of $8 \times 21$ points took $\approx 360$ hours using two processors 
\textit{Intel Xeon E5 2670 v3}
with 48 parallel threads in total.
The application of \EqRef{Eq:ElectronMAE} would increase this time by one or two orders of magnitude, not to mention the convergence problems highlighted above, which once again emphasizes the practical significance of the suggested approach.

\begin{figure}[htbp]
\includegraphics[width=0.35\textwidth]{\ContentPath{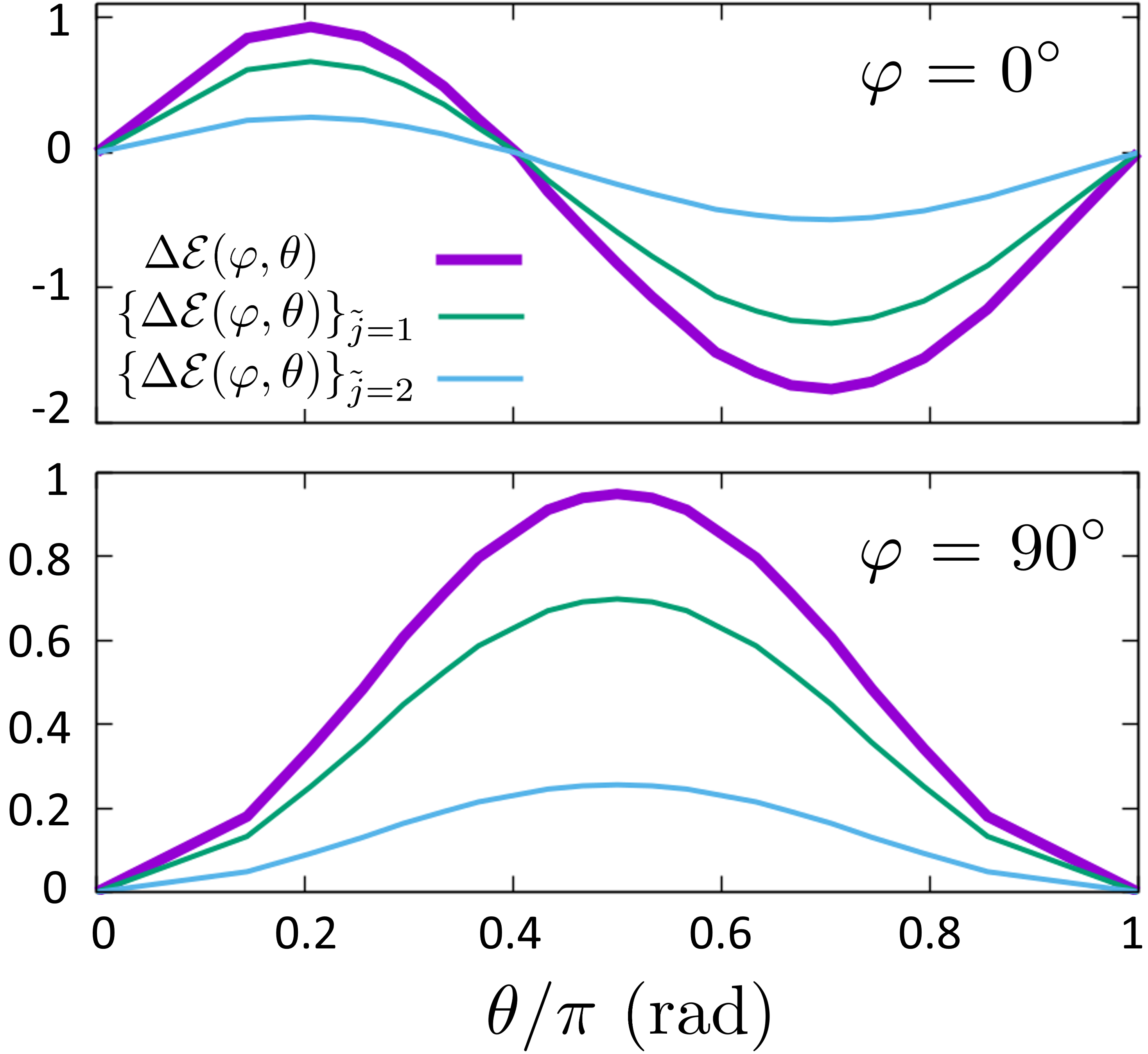}}
\caption{\label{fig:CoOMAE} 
MCA energy difference (meV) in CoO, obtained in the form of~\EqRef{Eq:EThetaPhiMinus00} for Co atom with local $(0, 0, 0)$~\AA~position in the unit cell.
\textit{(top)}    Angular profile with $\varphi = 0  \degree$.
\textit{(bottom)} Angular profile with $\varphi = 90 \degree$.
The contribution of the sublattice with this atom included is denoted as 
$\{ \deltaE(\varphi, \theta) \}_{ \sublattice{j}=1 }$. 
The case of another sublattice, which constitutes the ferromagnetic plane with the first one, is 
$\{ \deltaE(\varphi, \theta) \}_{ \sublattice{j}=2 }$.
The contributions of the rest two sublattices (with AFM orientation) are negligibly small.
The atom of each four sublattices in the unit cell with zero translation vector is given in the Appendix~\ref{AppendixA} as $\bm{r}_j$.
The illustration of the unit cell is given in Fig.~\ref{fig:CoOStructure} (\textit{bottom}).
}
\end{figure}

The resulting angular profiles of MCA energy difference obtained using \EqRef{Eq:FinalMAE} in the form of
\begin{equation}
\begin{split}
\deltaE(\varphi, \theta) &= \\  &\ElectronMAE(\varphi, \theta) -
                                 \ElectronMAE(\varphi = 0 \degree, \theta = 0 \degree) 
\end{split}
\label{Eq:EThetaPhiMinus00}
\end{equation}
for $\varphi = 0 \degree$ and $\varphi = 90 \degree$ are given in Fig.~\ref{fig:CoOMAE}. 
The Ref.~\cite{PhysRevB.74.155108} reports the curves of remarkable consistence with ours and traces the nature of them as proportional to the angular dependence profile of the orbital magnetic moment on the Co atom. 
Authors also performed the fitting of the MAE energy difference angular profile by trigonometric functions, designed for rhombohedral or monoclinic symmetry~\cite{Skomski}. It results in estimation of the easy axis direction as $[\bar{1}~0~7.2]$. However, the experimentally measured directions appears in a variation from $[\bar{1}~0~7.1]$ to $[\bar{1}~0~2.8]$~\cite{PhysRev.110.1333, D_Herrmann-Ronzaud_1978, PhysRevB.64.052102}. Thus, the origins of the easy axis in CoO are worthy to be under consideration.

For this purpose we firstly map our MAE energy angular profile onto the spin model, \EqRef{Eq:Spin_Hamiltonian}. Thus found SIAT reads (in meV)
\begin{equation}
[ A_{i = 1} ]^{*} = 
\begin{pmatrix}
  -2.512  &  0        &  0.567  \\
       0  &  -1.725   &  0      \\
   0.567  &  0        &  -2.147 
\end{pmatrix}
\, .
\label{Eq:CoOSIAT}
\end{equation}
Minimization of the \EqRef{Eq:Spin_Hamiltonian} with respect to $(e_x,~e_y,~e_z)$ yields $[\bar{1}~0~0.75] = [\bar{4}~0~3]$ for the easy axis direction. Equality $e_y = 0$ was anticipated directly from the rigorous symmetry of MAE energy angular profile:
$\ElectronMAE(e_x,~e_y,~e_z) = \ElectronMAE(e_x,~-e_y,~e_z)$,
provided by our local coordinate system adjustment.
One can readily mention that this axis lies close to the ferromagnetic plane (101) of Co atoms (the angle between this plane and the axis is $\approx8~\degree$).
It leads to hypothesis that the nature of this axis resides in the plane (101).

To provide an verification we employ an remarkable feature of \EqRef{Eq:FinalMAE}, that allows one to directly decompose the MCA energy value onto individual contributions of the sublattices:
\begin{equation}
\deltaE(\varphi, \theta) = 
   \sum_{ \sublattice{j} } 
      \{ \deltaE(\varphi, \theta) \}_{ \sublattice{j} }   \, .
\label{Eq:SublatticesDecomposition}
\end{equation}
We uncover a dramatic difference in proportions of these contributions. 
Two sublattices, which constitute the ferromagnetic plane with the Co atom $(0, 0, 0)$~\AA~included, appear with dominating role in developing of the whole MCA energy angular profile (Fig.~\ref{fig:CoOMAE}). Whereas contributions of the rest two sublattices, which describe the AFM related plane-plane interaction, are revealed nearly negligible ($\sim 0.1~\mu$eV). Thus we can reliably state our assumption reasonable and hence (101) planar physics drives the easy axis formation in CoO. 

It opens up a new perspectives towards experimental control of the MCA by applying a directional pressure along (101). Taking into account, that the character of the monoclinic distortion is known to be significantly sensible to hydrostatic pressure~\cite{PhysRevB.86.115134, PhysRevB.79.155123}, one could expect non-trivial influence on the magnetism of CoO in the proposed conditions.

In Fig.~\ref{fig:CoOMAE} we also find instructive to consider the difference between sublattices of the ferromagnetic plane. 
The sublattice with the Co atom $(0, 0, 0)$~\AA~included naturally appears special, because incorporates all purely on-site physics of this atom. To contrast it we should assume that these two sublattices by means of atom-lattice interaction contribute similarly to $\deltaE(\varphi, \theta)$. In this case the estimation of on-site effects impact could be performed as
\begin{equation}
\deltaOnSite(\varphi, \theta) =
\frac
{   |  \{ \deltaE(\varphi, \theta) \}_{ \sublattice{j} = 1 }
                                 -
       \{ \deltaE(\varphi, \theta) \}_{ \sublattice{j} = 2 }    |    }
{       | \deltaE(\varphi, \theta) |  }
  \, .
\label{Eq:OnSitePhysics}
\end{equation}
Thus we obtain 
for the maximum of $\varphi = 0  \degree$ profile, 
for the minimum of $\varphi = 0  \degree$ profile and
for the maximum of $\varphi = 90 \degree$ profile 
(Fig.~\ref{fig:CoOMAE}) 
$44.3 \%$,
$43.8 \%$ and 
$46.7 \%$, correspondingly.
Such uniformity in developing of the whole profile clearly points out the complicated nature of MCA in CoO, where the intra-atomic effects stand in line with spin-lattice interplay.


\section{Conclusion}

As a result of our research, we found out that the employment of electronic Green's functions with $\bm{k}$ resolution actually led to significantly more stable results of MCA energy for metallic systems. Testing on the effective model of individual \textit{d}-atom showed full agreement with the general physical expectations. The study of simple one-dimensional model with contrasted metallicity showed that the expression for MCA energy, based on inter-site Green's functions, has an asymptotic behavior as the maximal distance between atoms to be taken into account grows. The values to be converged to were successfully estimated using suggested approach, confirming its fruitfulness.

The \textit{ab initio} study of cobalt monoxide, empowered by the approach, has lifted the veil from the physical mechanisms, underlying the forming of the MCA picture. The decomposition of MCA energy profile into the contributions of four Co sublattices proved the ferromagnetic Co plane dominant in generating of the easy axis and enabled one to qualitatively distinguish itinerant mechanisms from the intra-atomic Co effects.

Thus, the approach suggested in this work was shown effective for studying the effects of magnetocrystalline anisotropy in both model and real materials with a metallic electronic structure.



\section{Acknowledgments}

We thank Vladimir V. Mazurenko for fruitful discussions.
The work is supported by the
Russian Science Foundation, 
Grant No. 23-72-01003,
https://rscf.ru/project/23-72-01003/


\appendix
\section{\textit{Ab initio} calculations of CoO}
\label{AppendixA}

We performed the first-principles calculations of CoO electronic structure within density functional theory (DFT)~\cite{PhysRev.140.A1133}. For this purpose we employ the generalized gradient approximation with an effective Coulomb repulsion on the Co atoms (GGA\textit{+U}) 
and the Perdew-Burke-Ernzerhof (PBE) exchange-correlation functional~\cite{PhysRevLett.77.3865}.
The GGA\textit{+U} calculations are based on the scheme of Dudarev~\textit{et~al.}~\cite{PhysRevB.57.1505} and we perform it using the Quantum-Espresso simulation package~\cite{Giannozzi_2009}.

The basic parameters of the simulation are following:
\begin{itemize}
\item The energy cutoff of the plane wave basis construction is set
to 330~eV;
\item The energy convergence criterion is $10^{-6}$~eV;
\item The $20 \times 20 \times 20$ Monkhorst-Pack grid was employed to carry out integration over the Brillouin zone;
\item The lattice vectors are:
$\bm{a} = (6.07, 0, 0)$~\AA; 
$\bm{b}~=~(0, 2.96, 0)$~\AA;
$\bm{c} = (3.07, 0, 4.23)$~\AA;
\item The local positions of four Co atoms in the unit cell are:
$\bm{r}_1 = (0, 0, 0)$~\AA;
$\bm{r}_2 = (4.57, 1.48, 2.11)$~\AA;
$\bm{r}_3 = (3.03, 0, 0)$~\AA;
$\bm{r}_4 = (1.54, 1.48, 2.11)$~\AA;
\item An effective Coulomb repulsion parameter is $U~=~4$~eV, in consistency with Ref.~\cite{PhysRevB.86.115134}.
\end{itemize}

\begin{figure}[htbp]
\includegraphics[width=0.3\textwidth]{\ContentPath{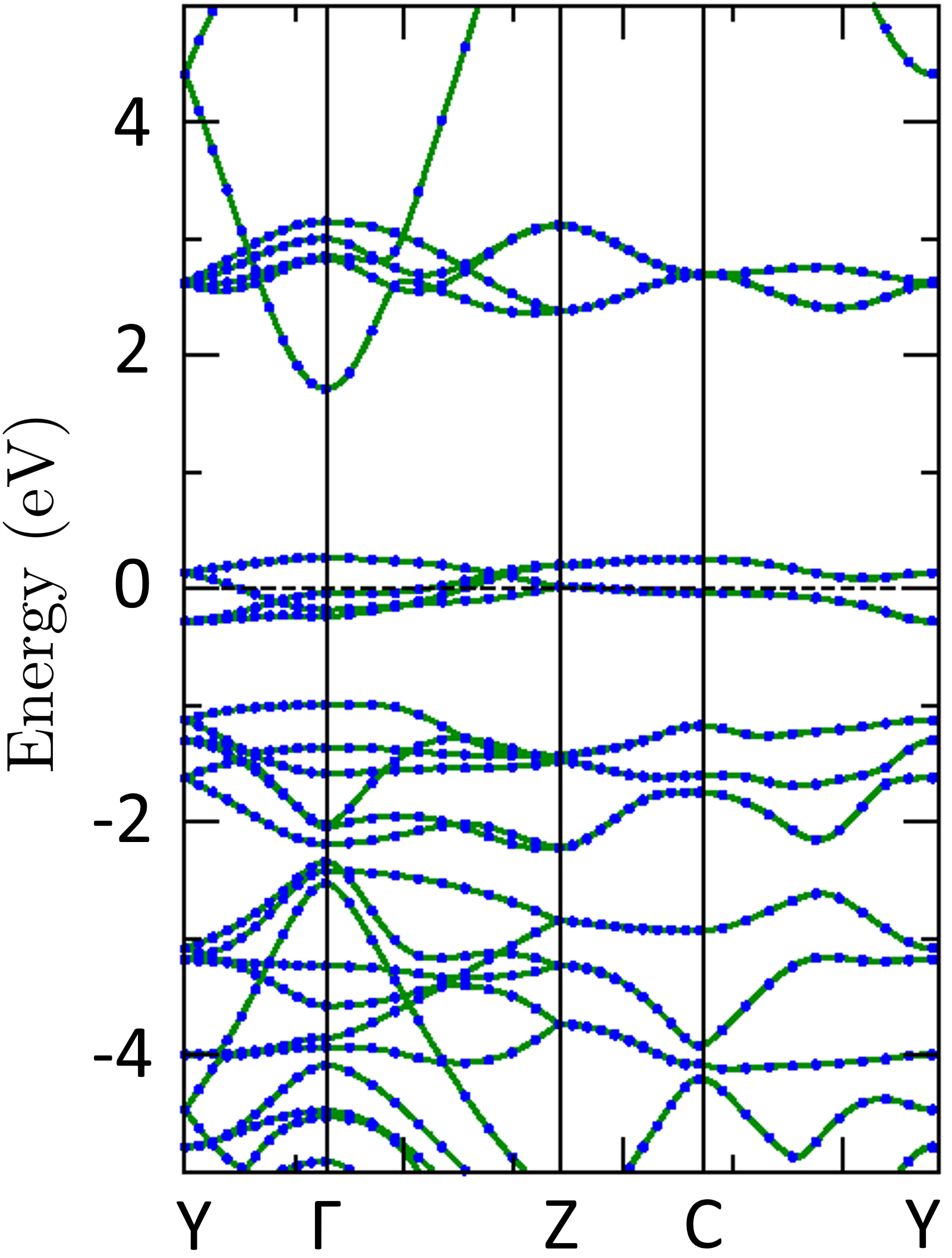}}
\caption{\label{fig:CoOBands} 
The band structure of CoO, obtained from GGA\textit{+U} calculations (green solid line) 
and from low-energy model (blue points). 
The high symmetry points are
 $\mathrm{Y} (0.5, 0, 0)$,
 $\Gamma(0, 0, 0)$,
 $\mathrm{Z}(0.5, 0.5,0)$,
 $\mathrm{C}(0.5, 0.5,0)$. 
 The Fermi level is zero.
}
\end{figure}

Fig.~\ref{fig:CoOBands} gives the resulting band structure of CoO, which is in excellent agreement with that presented in previous theoretical studies~\cite{i9061050}. 

To construct the low-energy model of magnetoactive Co $3d$-shell we project obtained from GGA\textit{+U} wave functions onto maximally localized Wannier functions
~\cite{PhysRevB.56.12847, PhysRevB.65.035109, MOSTOFI2008685}, 
describing along with the $3d$ states also $4s,~4p$ states for cobalt and $2s,~2p$ states for oxygen, because of the essential entanglement of corresponding bands. 
As it is seen from Fig.~\ref{fig:CoOBands}, the resulting model perfectly reproduces DFT band structure in wide energy range.

\bibliography{\ContentPath{Biblio}}


\end{document}